\documentclass[sigconf]{acmart}

\AtBeginDocument{%
  }

\begin{document}

\title{DLRREC: Denoising Latent Representations via Multi-Modal Knowledge Fusion in Deep Recommender Systems}

\author{Jiahao Tian}
\authornote{Both authors contributed equally to this research.}
\affiliation{%
  \institution{Georgia Institute of Technology}
  \city{Atlanta}
  \state{Georgia}
  \country{USA}
}
\email{jtian83@gatech.edu}

\author{Zhenkai Wang}
\authornotemark[1]   
\affiliation{%
  \institution{The University of Texas at Austin}
  \city{Austin}
  \state{Texas}
  \country{USA}
}
\email{kay.zhenkai.wang@utexas.edu}


\begin{abstract}
Modern recommender systems struggle to effectively utilize the rich, yet high-dimensional and noisy, multi-modal features generated by Large Language Models (LLMs). Treating these features as static inputs decouples them from the core recommendation task. We address this limitation with a novel framework built on a key insight: deeply fusing multi-modal and collaborative knowledge for representation denoising. Our unified architecture introduces two primary technical innovations. First, we integrate dimensionality reduction directly into the recommendation model, enabling end-to-end co-training that makes the reduction process aware of the final ranking objective. Second, we introduce a contrastive learning objective that explicitly incorporate the collaborative filtering signal into the latent space. This synergistic process refines raw LLM embeddings, filtering noise while amplifying task-relevant signals. Extensive experiments confirm our method's superior discriminative power, proving that this integrated fusion and denoising strategy is critical for achieving state-of-the-art performance. Our work provides a foundational paradigm for effectively harnessing LLMs in recommender systems.
\end{abstract}



\maketitle

\section{Introduction}
Recommender Systems (RS) have emerged as a vital element in today's digital ecosystems, significantly contributing to the personalization of user experiences across platforms such as e-commerce, streaming services, and content delivery systems \cite{guo2017deepfm,zhou2018deep,yin2023heterogeneous}. These systems strive to recommend items that match users' preferences, thereby boosting engagement and satisfaction. The origins of RS often lie in collaborative filtering techniques, which analyze user-item interaction data to uncover patterns and generate suggestions. As the volume and complexity of user data have increased, these systems have evolved to adopt more advanced methodologies, including content-based filtering, hybrid models, and context-aware recommendation strategies.

The development of machine learning, particularly deep learning, has revolutionized nearly every field\cite{li2025chatmotion, dan2024image,tian2025time,tian2024deepcensored,zhang2025mamnet}, including recommender systems \cite{tian2024mmrec,ding2024semantic,zhao2025efficient}. These systems now benefit from large-scale models capable of leveraging vast amounts of data to extract complex relationships and patterns. Deep learning techniques, such as those employed in modern Deep Learning Recommendation Models (DLRM) frameworks and Graph Neural Networks (GNNs), have been employed to enhance the accuracy and robustness of RS. These models benefit from their ability to automatically learn feature representations from diverse raw data, eliminating the need for extensive manual feature engineering and improving predictive performance.

The remarkable capabilities of Large Language Models (LLMs) like GPT-4 in processing nuanced, multi-modal information from text and images have opened new frontiers for recommender systems (RS). However, a prevailing paradigm for leveraging LLMs involves their use as a disconnected pre-processing tool—a powerful "feature extractor" where high-dimensional embeddings are generated offline and later fed into a separate recommendation model. This decoupled, two-step pipeline is fundamentally limited, as the feature extraction is not guided by the final recommendation objective, and it often fails to incorporate essential collaborative filtering signals.

In contrast to prior methods that treat large language models as separate, upstream feature extractors, this paper introduces a unified, end-to-end framework. Our architecture cohesively integrates dimensionality reduction directly with the primary recommendation task. By co-training these components, we ensure that the learned low-dimensional representations are optimized specifically for the ranking objective, rather than being a generic byproduct of a separate process.
A key innovation of our model is the enrichment of these embeddings with collaborative filtering signals through a novel, multi-relational contrastive learning objective. This infusion of user-user and item-item similarities provides a richer, more robust basis for representation learning. Furthermore, a significant advantage of our design is its model-agnostic nature, allowing for straightforward integration into a wide variety of existing recommendation architectures. This holistic approach yields compact, robust, and highly discriminative representations, ultimately leading to superior recommendation performance.to demonstrably superior performance over conventional two-step methodologies. Main contributions of this paper include:
\begin{itemize}
    \item \textbf{A co-trained dimensionality reduction mechanism}: We propose integrating dimensionality reduction directly into the recommender framework, enabling end-to-end co-training with the primary recommendation objective. This ensures the generation of highly task-relevant latent representations from LLM-derived multi-modal features.
    \item \textbf{A contrastive learning objective for enhanced collaborative signal integration}: We introduce an auxiliary task that leverages pre-computed user-user and item-item similarities (via SWING and InfoNCE loss) to explicitly model these relationships, thereby enriching representation quality and robustness of the resulting multi-modal representation in the reduced embedding space.
    \item \textbf{Demonstration of enhanced LLM-based multi-modal recommendation}: We empirically show how these architectural and methodological innovations lead to improved recommendation performance by fostering more effective and robust user and item representations, highlighting the utility of our proposed techniques.
    \item \textbf{An open-sourced\footnote{Our code is open-sourced at \url{https://github.com/consistentJake/LLM\_multimodal}.}, readily adaptable module}: We provide a publicly available implementation of our contrastive learning-based dimensionality reduction module, designed for easy integration into existing or new recommender frameworks, facilitating further research and application.
\end{itemize}

The structure of the paper is as follows: Section II introduces related work concerning deep learning in recommender systems and the emerging role of LLMs. Section III presents our methodology, detailing its key components including the foundational LLM-based knowledge extraction, the integrated co-trained dimension reduction, and the multi-relational contrastive learning objectives. Section IV details the experimental setup, datasets, and analysis of results. Finally, Section V provides concluding remarks and discusses future work.

\section{Related Work}
\subsection{Large Language Models in Recommender Systems}
LLM has transformed different areas since its inception \cite{wang2025digital,ji2025mitigating,huang2025enhancing}. The integration of LLM into recommender systems has become increasingly popular ~\cite{ji2025reason,gptsignal}. Recent efforts have explored a variety of applications, including interactive recommendation interfaces, prompting strategies, and hybrid models~\cite{hou2022towards,wang2022towards}.

For instance, \cite{gao2023chat} introduced an LLM-driven conversational framework to enable multi-turn recommendations, improving both engagement and explainability. Another approach by \cite{wang2023generative} utilizes user interaction data and natural language feedback processed through an instructor module to guide LLM-based recommendation generation. In a systematic study, \cite{dai2023uncovering} assessed the effectiveness of off-the-shelf LLMs for recommendation tasks using point-wise, pair-wise, and list-wise evaluation metrics.

\subsection{Constrastive Learning in Recommender Systems}
Recent research in recommendation systems has increasingly leveraged contrastive learning frameworks to learn more effective user and item representations from the use of this semi-supervised learning approach. For instance, the construction of positive pairs from temporally adjacent items in a user's history or augmented subsequences showed promising results in sequential recommendation \cite{zhou2020s3}. The constrastive loss was demonstrated to be useful by leveraging positive/negative pairs defined within the interaction graph in the context of graph-based collaborative filtering models \cite{wu2021self}. The application of contrastive loss functions helps in creating a more discriminative embedding space and has demonstrated potential in alleviating issues like data sparsity and the cold-start problem \cite{oord2018representation}.

\subsection{Traditional and Neural Approaches in Recommendation}

Traditional recommender systems were originally built without deep learning. A survey of over 100 such methods prior to 2017 is presented in~\cite{app7121211}. Modern systems can be classified broadly into personalized\cite{wu2022knowledge} and group-based\cite{zan2021uda} categories.

Collaborative Filtering (CF) has remained a cornerstone technique. Memory-based CF approaches such as those in~\cite{chen2019collaborative,barkan2021anchor} utilize vector representations to model user-item interactions. The emergence of graph neural networks (GNNs) has further enhanced model-based CF. Models like GraphSAGE~\cite{hamilton2017inductive} have demonstrated notable improvements across domains such as music \cite{doh2025talkplay}, books, and Points of Interest (POI)\cite{kumar2019predicting,sun2019multi}. 

In addition, review texts have been extensively explored as supplementary signals. \cite{chin2018anr} proposed an aspect-based neural recommender (ANR) that selectively attends to informative review segments. \cite{li2019capsule} employed capsule networks to capture fine-grained opinion expressions. \cite{wu2019npa} developed a dual-encoder CNN-based framework for modeling user preferences via historical interactions with news articles.

\subsection{LLMs for External Knowledge Generation in Recommendation}
LLMs have demonstrated strong reasoning skills. Notable techniques include few-shot prompting~\cite{gpt3} and Chain-of-Thought reasoning~\cite{cot}, which improve performance in arithmetic and commonsense reasoning benchmarks. The capabilities of LLM shown has laid a solid foundation for its use in recommender systems.

For instance, KAR~\cite{xi2024towards} augments movie and book recommendation by generating factual side information. LLMRec~\cite{wei2024llmrec} improves user/item profiles and tackles data sparsity by constructing pseudo-supervised training data. ONCE~\cite{liu2024once} uses prompting techniques to generate content features for new items and refine user profiling. These methods rely on the generative capabilities of LLMs to enrich traditional recommendation pipelines.

In contrast, our approach takes a more conservative path by extracting recommendation-relevant knowledge from existing external sources rather than relying solely on generated content. This design choice aims to reduce hallucination risks often associated with LLMs~\cite{zhao2023survey}, thereby enhancing the credibility and robustness of the incorporated knowledge.

\section{Methodology}
This research advances recommendation systems by introducing significant enhancements to a Deep Learning Recommendation Model (DLRM) framework \cite{naumov2019deep}. While retaining the core DLRM structure that processes diverse input features (sparse categorical attributes, dense numerical data, and high-dimensional embeddings representing users, items, and associated content), this paper details two key methodological innovations: the integration and co-training of a dimension reduction module, and the incorporation of collaborative filtering principles via a contrastive learning objective.

\subsection{Foundational DLRM Framework}
At a high level, DLRM processes sparse categorical features through embedding layers and handles dense numerical features directly through an MLP. A key aspect is its explicit feature interaction stage (e.g., dot products between embeddings) followed by a final MLP prediction layer to generate the recommendation score.
To enrich the inputs for such models, a common strategy involves leveraging Large Language Models (LLMs) to extract and fuse knowledge from multi-modal sources. This can be achieved in several ways:
\begin{itemize}
    \item Textual Content: LLMs can be employed to read and summarize collections of user reviews. The resulting summary, capturing the core essence, can then be embedded and used as a refined textual representation, potentially reducing noise compared to aggregating all review embeddings.
    \item Visual Content: LLMs capable of multi-modal processing can interpret associated images, generating textual descriptions. Embedding these descriptions translates visual information into the same semantic space as text features, potentially unifying multi-modal signals.
    \item Feature Extraction: LLMs can also perform direct feature extraction, such as inferring pricing levels (treated as dense features) or classifying items into categories (treated as sparse features) from unstructured text.
\end{itemize}
These LLM-generated outputs (summary embeddings, image description embeddings, extracted dense/sparse features) would then be integrated into the appropriate pathways of the DLRM. However, content embeddings derived from such methods often possess high dimensionality. Consequently, a typical approach to manage this complexity involves employing a separate, upstream dimension reduction step (e.g., an MLP) applied to these high-dimensional embeddings before the main DLRM feature interaction stage. This reduction module is often optimized independently of the final ranking task. This two-phase approach with individual framework for dimension reduction often leads to sub-optimal performance due to the loss of information in the dimension reduction phase.

\subsection{Integrated Co-trained Dimension Reduction}

A primary contribution of our methodology is the direct integration of the dimensionality reduction process within the main DLRM architecture, enabling it to be trained jointly with the overall recommendation task. In contrast to approaches that rely on a separate pre-processing step, our model ingests the raw, high-dimensional embeddings of the user, item, and any associated content for each training instance. These embeddings are then fed into dedicated neural network layers embedded within the DLRM.

During the model's forward pass, these integrated layers perform the dimensionality reduction, producing lower-dimensional latent representations of the initial high-dimensional inputs. These learned, compressed embeddings are subsequently passed to downstream components, such as feature interaction modules, where they are combined with other processed sparse and dense features to generate the final prediction.
Crucially, the parameters of this embedded dimension reduction component are optimized simultaneously with all other model parameters—including embedding tables, interaction layers, and the prediction head—through end-to-end backpropagation. This co-training is driven by the final recommendation objective's loss function, such as predicting user interaction probability. This approach ensures that the dimensionality reduction is specifically tailored to the ranking task, learning to retain and emphasize the features most salient for accurate recommendations. This yields more effective and task-relevant latent representations than what is typically achievable with a separately optimized, upstream reduction step.

\subsection{Collaborative Filtering via Multi-Relational Contrastive Learning}
To enhance the quality of the learned representations, we introduce a multi-relational contrastive learning objective. This approach addresses the common challenges of data sparsity and noise inherent in user-item interaction logs by incorporating stronger collaborative filtering signals into the representation learning process. Unlike conventional methods that rely solely on user-item interactions, our framework explicitly models multiple types of relationships by leveraging item-item and user-user similarities alongside the primary user-item interaction data.

To establish these supplemental relationships, we first compute similarity scores between users and between items using the SWING (Similarity Weighted by Inverse Neighborhood frequency) algorithm. SWING is a graph-based method well-suited for recommendation data \cite{luo2024trawl,zhang2024notellm}, as it identifies similarities by analyzing the higher-order structure of the user-item interaction graph. High similarity between two entities (e.g., two users) indicates a strong relationship based on shared interaction patterns with common items, considering the overlap characteristics of those intermediaries. This pre-computed similarity information serves as a robust foundation for our contrastive learning task, guiding the selection of positive samples to ensure that the learned embeddings are informed by a richer, multi-relational understanding of the collaborative filtering domain.

We first establish similarity metrics between users and between items. We utilize the SWING (Similarity Weighted by Inverse Neighborhood frequency) algorithm, a graph-based approach suitable for recommendation data. Conceptually, SWING identifies similar users (or items) by analyzing the structure of the user-item interaction graph. High similarity between two users (or two items) suggests they are strongly related based on shared interaction patterns with common items (or users), considering the overlap characteristics of those intermediary entities. This pre-computed similarity information is used to guide the selection of positive samples in the contrastive learning phase.

    We employ the InfoNCE \cite{gutmann2010noise} loss function to implement the contrastive learning objective for both user-user ($L_{uu}$) and item-item ($L_{ii}$) similarities. The goal is to pull embeddings of similar entities closer together in the latent space while pushing dissimilar entities apart.For the user-user contrastive loss , considering a batch of user embeddings, the loss for an anchor user j is defined as:
    \begin{equation}
    L_{uu}^{(j)} = -\log \frac{\exp(u_j \cdot u_{j}^{p}/\tau)}{\exp(u_j \cdot u_{j}^{p}/\tau) + \sum_{k \in \mathcal{N}_j} \exp(u_j \cdot u_{j,k}^{n}/\tau)}
\end{equation}

Where:
\begin{itemize}
    \item $u_j$ is the dimension-reduced embedding for user $j$
    \item $u_{j}^{p}$: is the embedding of a 'positive' user sample for user $j$. This positive sample is identified using the pre-computed SWING similarity scores (i.e., a user highly similar to user $j$).
    \item $\mathcal{N}_j$: is the set of 'negative' user samples for user $j$. These negative samples typically consist of other users  randomly selected from the corpus, assumed to be dissimilar to user $j$.
    \item $\tau$ is the temperature hyperparameter, controlling the sharpness of the distribution and influencing the separation between positive and negative pairs.
\end{itemize}

The total user-user loss $L_{uu}$ is the average of $L_{uu}^{(j)}$ over all anchor users in the batch. An analogous loss term, $L{ii}$, is computed for item-item similarity using item embeddings and the corresponding SWING-derived item similarities to select positive item samples.

The overall training objective combines a primary recommendation loss with auxiliary contrastive losses designed to leverage collaborative signals. This composite loss guides the model to learn representations that are effective for both predicting user-item interactions and capturing underlying user-user and item-item similarities.

\begin{equation}
    L = L_{rec} + w_{1}L_{ii} + w_{2}L_{uu}
\end{equation}

\begin{itemize}
    \item $Lrec$ is the loss associated with the main recommendation task.
    \item $Lii$ and $Luu$ are the item-item and user-user InfoNCE contrastive losses, respectively.
    \item $w1$ and $w2$ are scalar hyperparameters that balance the influence of the collaborative contrastive signals relative to the primary recommendation objective.
\end{itemize}

\begin{figure*}[hbt!]
    \centering
    \includegraphics[width=\linewidth, height=0.3\textheight]{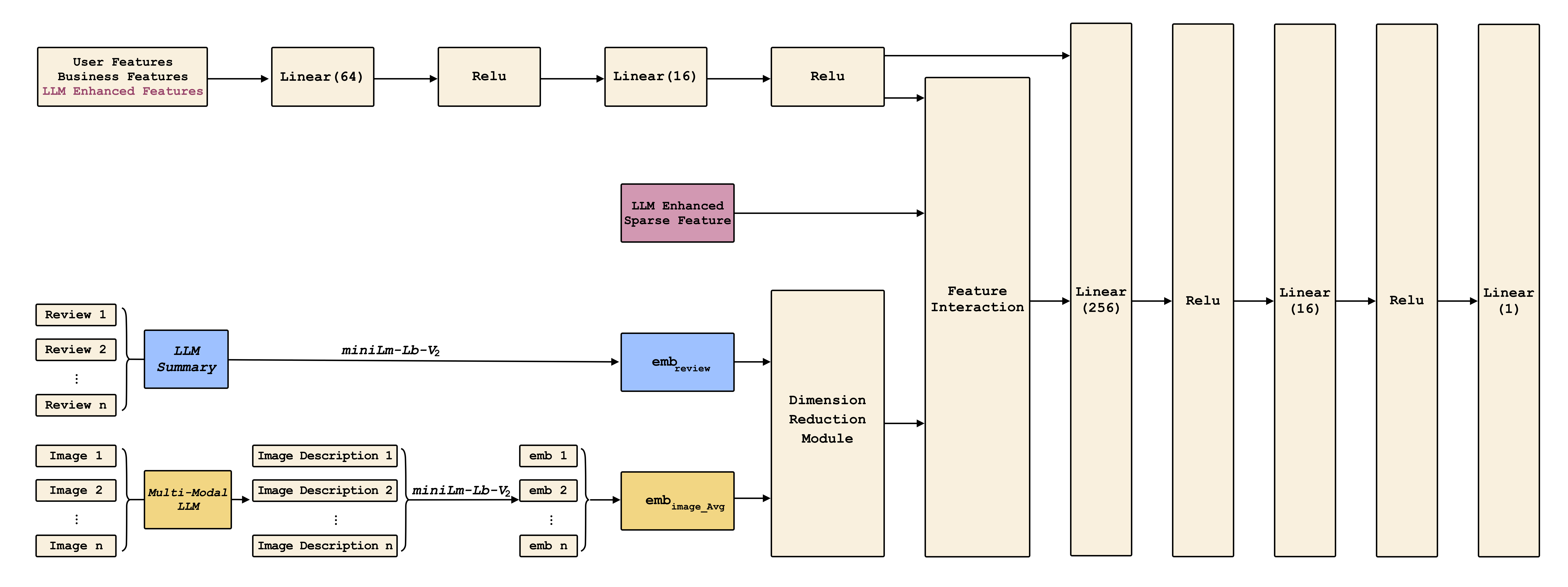}
    \caption{Proposed \textbf{Unified} Model Architecture with \textbf{Contrastive-Learning Based} Dimension Reduction Module}
    \label{fig:model_arch}
\end{figure*}

\section{Experiment Setup}
\subsection{Data Source}
In this study, we used a comprehensive data set tailored for restaurant reviews analysis. This data is published in Kaggle\footnote{\url{https://www.kaggle.com/}} and it is collected from Google reviews~\cite{googlerestaurantreviews}. This data set comprises user-generated text reviews and image reviews for various restaurants.The dataset comprises user-generated reviews for a diverse range of restaurants. Each record contains the following components:
\begin{itemize}
\item A unique user identifier,
\item The business identifier corresponding to the reviewed restaurant,
\item A review rating on a scale from 1 to 5 (with ratings $\geq$ 4 considered positive),
\item The full text of the user's review,
\item Images uploaded by the user that are linked to the review.
\end{itemize}

\subsection{Parameter and Configuration}
To evaluate the model's performance under different conditions, we experiment with various dropout rates, different weighted loss functions, and baseline vs proposed models.

\begin{itemize}
    \item \textbf{Dropout:} Dropout is a regularization technique that helps prevent overfitting by randomly setting a fraction of input units to zero during training. We experiment with dropout rates of [0.1, 0.3, 0.5].

    \item \textbf{Weighted Loss:} This dataset is highly imbalanced, with only 1/8 of the datapoints labeled as false and the rest as positive. To balance the contribution of each class to the loss function, we experiment with different class weightings.
\end{itemize}

Each training run starts with a learning rate of 0.01, using Adaptive Moment Estimation (Adam) for adaptive learning rate adjustment. In each epoch, we evaluate model performance on both the training and test sets. The best model is updated when a better false positive rate is achieved on the test set. Early stopping is applied starting after 300 epochs. Training terminates if there is no continuous improvement in the false positive rate over the last 50 epochs. Each parameter configuration is evaluated five times.

\subsection{Data Pre-processing}

Our baseline model utilizes dense and embedding features to predict the review rating. To avoid data leakage, the current review is excluded during feature construction.

The baseline model includes new dense and sparse features:

\begin{itemize}
    \item \textbf{Dense Feature:}
    \begin{itemize}
        \item \textit{Price Tag:} Using GPT-3.5-turbo-1106, we prompt reviews with ``Can you tell me if the price is overpriced, fair price, low price from reviews for this restaurant? Give me just the category.'' After post-processing, a price category (fair, overpriced, cheap, or none) is assigned to each restaurant.
    \end{itemize}

    \item \textbf{Sparse Feature:}
    \begin{itemize}
        \item \textit{Restaurant Category:} Reviews are processed using the prompt ``Can you tell me what kind of restaurant this is from these reviews for the restaurant? Return me in this format: 'type'.'' This results in a list of up to 11 subtypes per restaurant. The output is padded to 11 entries, using index 179 for padding (out of 180 distinct embedding values).
    \end{itemize}

    \item \textbf{Embedding Text Features with LLM:}
    \begin{itemize}
        \item \textit{User and Business Summary Embeddings:} We use GPT-based prompts (shown in Fig.~\ref{fig:prompts}) to summarize all reviews written by a user or received by a business. These summaries are embedded into 384-dimensional vectors. To better capture contrastive characteristics, we use separate projection models for users and businesses. The user summary embedding is passed through a user-specific projection model, while the business summary embedding is passed through a business-specific projection model. Both projection models are trained using contrastive learning and reduce the embeddings from 384 to 32 dimensions.
    \end{itemize}



\begin{figure*}[t!]
    \centering
\includegraphics[width=\linewidth]{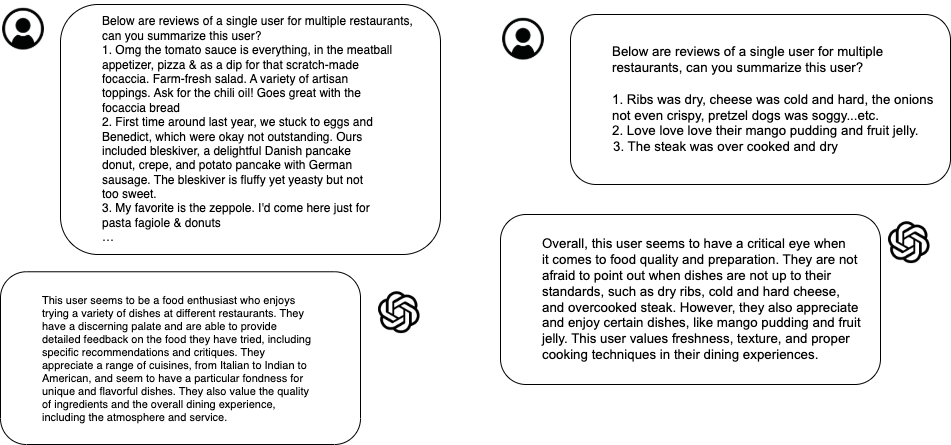}
    \caption{Examples of leveraging to summarize user reviews for same restaurant. The same prompting techniques were used as in ~\cite{tian2024mmrec}}
    \label{fig:prompts}
\end{figure*}

    \item \textbf{Embedding Image Features with Image Captioning:}
    \begin{itemize}
        \item Using BLIP-2 for unconditional image captioning, each image is converted into a descriptive sentence, which is embedded and average-pooled into a comprehensive image representation. This vector is also passed through a projection model and converted into a 32-dimensional vector.
    \end{itemize}
\end{itemize}

\begin{table*}[t!]
\centering
\small
\begin{tabular}{|l|l|c|c|c|c|}
\hline
\textbf{Model} & \textbf{Loss} & \textbf{Acc.} & \textbf{FP} & \textbf{V0 Acc.\footnotemark[1]} & \textbf{V0 FP} \\ \hline
Text Only & BCE + Contr. & 99.3 & 0.36 & n/a & n/a \\ \hline
Text Only & BCE & 97.70 & 1.07 & 79.10 & 19.10 \\ \hline
Image Only & BCE & 89.10 & 16.20 & 57.90 & 28.30 \\ \hline
Text + Image & BCE + Contr. & 99.70 & 0.15 & n/a & n/a \\ \hline
Text + Image & BCE & 98.46 & 1.20 & 83.50 & 18.20 \\ \hline
\end{tabular}
\caption{Model Performance Comparison: best accuracy and false positive rate against test set with different experiment setting}
\label{tab:model_comparison}
\end{table*}

\footnotetext[1]{V0 model refers to the model in  \cite{tian2024mmrec}}
\subsection{Base Model}
We refer to the model proposed in \cite{tian2024mmrec} as the V0 model, which in sharp contrast with our single unified framework, adopts a two-step approach that trained MLP-based model to obtain multi-modal redpresentation in the reduced embedding space, followed by training a DLRM model on the reduced features. We believe that this two step approach could lead to sub-optimal performance due to the lack of user/item interaction information during dimension reduction phase. we build on DLRM by introducing key innovations: collaborative training, contrastive learning, and a unified architecture. Considering the theme shared between two works, we followed the data processing and feature engineering steps laid out in MMREC\cite{tian2024mmrec}. By directly comparing our proposed methods against the previous work that uses the same multi-modal information, we aim to demonstrate the effectiveness of the integration of dimension reduction into the recommender system and the use of contrastive learning module to fuse collaborative filtering signal into the recommendation task.

\subsection{Results and Discussion}

The results clearly demonstrate the advantages of the proposed framework. The new models significantly outperform the V0 baseline, with even BCE-only configurations showing substantial gains (e.g., ~18\% higher accuracy, ~18\% lower FP rate compared to V0 counterparts in Table \ref{tab:model_comparison}). This highlights the benefits of the co-trained dimension reduction module learning task-specific representations.

The addition of the multi-relational contrastive loss yielded further, critical improvements. Accuracy increased consistently (reaching 99.7\% for Text + Image), and false positive rates plummeted – notably from 1.2\% down to 0.15\% for the best model. This dramatic FP rate reduction strongly indicates that the contrastive loss fosters more discriminative embeddings. By structuring the latent space according to collaborative similarities (user-user, item-item via SWING), the model becomes better at separating relevant items from irrelevant ones, directly enhancing ranking precision.

Furthermore, the high overall accuracy suggests the contrastive objective helps create richer, more robust representations. Incorporating these auxiliary collaborative signals improves generalization beyond learning from direct interactions alone.

\section{Conclusion}
In this paper, we addressed the critical challenge of enhancing deep learning-based recommender systems through representation learning from multi-modal data. Our primary contribution lies in a novel framework that uniquely integrates dimensionality reduction directly into the recommendation model, enabling end-to-end co-training, and introduces a multi-relational contrastive learning objective. This approach marks a significant departure from conventional methods that typically treat dimensionality reduction as an isolated pre-processing step and often overlook the rich, explicit user-user and item-item similarity signals for representation refinement.

By synergistically combining co-training with a multi-relational contrastive loss that leverages explicit collaborative signals (e.g., derived from SWING and InfoNCE), our framework demonstrably generates richer, more robust, and more discriminative embeddings from LLM-processed multi-modal inputs. Crucially, our method for embedding collaborative filtering principles directly within the dimensionality reduction process offers a versatile solution, readily applicable to a wide array of recommendation tasks via our publicly available open-sourced module.

Future research avenues include exploring alternative similarity metrics beyond SWING, investigating different formulations for contrastive learning, and extending the application of this integrated framework to diverse recommendation domains and evolving multi-modal data types. Ultimately, our work provides a foundational step towards more holistic and contextually aware representation learning in next-generation recommender systems.

\bibliographystyle{ACM-Reference-Format}
\bibliography{bibliography}

\end{document}